\documentclass[slac_one]{revtex4}
\usepackage{latexsym,epsfig,graphicx}
\usepackage[dvips]{color}
\usepackage{fancyhdr}
\newcommand{\EPEM}{\mbox{$e^+e^{-}$}}
\newcommand{\EMEM}{\mbox{$e^-e^-$}}
\newcommand{\GG}{\mbox{$\gamma\gamma$}}
\newcommand{\GE}{\mbox{$\gamma e$}}

\newcommand{\TEV}{\mbox{TeV}}

\newcommand{\CM}{\mbox{cm}}

\newcommand{\MKM}{\mbox{$\mu$m}}

\newcommand{\ENX}{\mbox{$\epsilon_{nx}$}}
\newcommand{\ENY}{\mbox{$\epsilon_{ny}$}}

\newcommand{\be}{\begin{equation}}
\newcommand{\ee}{\end{equation}}
\newcommand{\bc}{\begin{center}}
\newcommand{\ec}{\end{center}}
\newcommand{\bi}{\begin{itemize}}
\newcommand{\ei}{\end{itemize}}

\begin{document}

\title{{\small{2005 International Linear Collider Workshop - Stanford,
U.S.A.}}\\ 
\vspace{12pt} Crossing angle at the photon collider} 

%

\author{V.I. Telnov}
\affiliation{Institute of Nuclear Physics, 630090 Novosibirsk, Russia}

\begin{abstract}
   For removal of disrupted beams at the ILC linear \EPEM\ collider it
   is desirable to collide beams at some crossing angle.  An
   especially large crossing angle, of about 25 mrad, is necessary for
   the photon collider, where disrupted beams are softer and wider
   than in the \EPEM\ case.  Some complications arise due to the
   solenoidal magnetic field of the detector. Radiation of particles
   in this field can increase the vertical beam size, leading to a
   loss of the luminosity. In addition, in the presence of a detector
   magnetic field the beam trajectories are not flat. Moreover, in the
   case of \EMEM\ beams, the vertical collision angle is non-zero and
   larger than the beam diagonal angle $\sigma_y/\sigma_z$, and this
   should be corrected in some way.  In this paper, possible solutions
   of these problems are considered. It is shown that at the ILC one
   can have the interaction region compatible both with \EPEM\ and
   \GG, \GE\ modes of operation without loss of luminosity.
\end{abstract}

\maketitle
\thispagestyle{fancy}
\section{INTRODUCTION} %

   At photon colliders, high energy photon beams are produced by
   Compton scattering of laser photons off high energy electrons in
   linear colliders~\cite{GKST81,GKST83}. Due to multiple Compton
   scattering the beams after the $e\to\gamma$ conversion have a
   large energy spread:
   $E=(0.02$--$1)E_0$~\cite{TEL90,TEL95,TESLAgg,TESLATDR}. After the
   collision with the opposing electron beam,  low energy electrons
   acquire disruption angles $\theta_d \sim$ 10--12 mrad (background
   from particles with larger angles is less than that from other
   unavoidable backgrounds)~\cite{TESLATDR}. Large energy and angular
   spreads of the beams present a problem of the removal of beams
   from the detector.

\begin{figure}[b]
\vspace{-0.5cm}
\epsfig{file=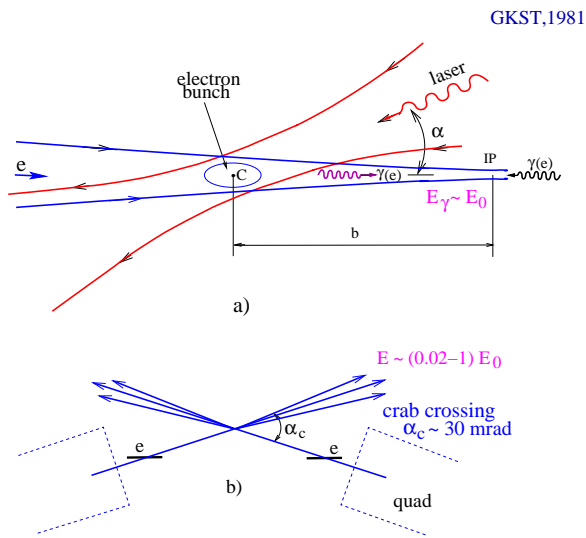,width=8cm,angle=0}
\caption{Scheme of the beam collisions at the photon collider}
\label{fig1}
\end{figure}

  The crab-crossing scheme of the beam collisions at \EPEM\ linear
  colliders~\cite{Palmer} presents a very nice solution to the above
  problems at the photon collider~\cite{TEL90}, Fig.~\ref{fig1}.
  Though the collision angle $\alpha_c$ is larger than the horizontal
  diagonal angle of the beam at the interaction point (IP),
  $\sigma_x/\sigma_z$, there is no loss of the luminosity because each
  beam is tilted using a special RF cavity) by the angle $\alpha_c/2$.
Due to this collision angle, the outgoing disrupted beams travel outside
the final quads.  The value of the crab--crossing angle is determined
by the disruption angles and by the final quad design (the diameter of the
quad and its distance from the IP):
\be
\alpha_c \sim
R_{quad}/L^* + \theta_d \\ \sim 6\,\CM\ / 400\,\CM\ +0.01 \sim 25 \; 
\mbox{mrad}.
\ee
For \EPEM\ collisions, $\alpha_c = 20$ mrad is one of the considered options.
It is very desirable to have the crossing compatible with both
collision modes, i.e. $\alpha_c \sim$ 20--25 mrads. 

There are several problem associalted with a large crab-crossing angle:
\bi
\item stability of the beam tilt (provided by the crab-cavity) with
an accuracy better than $(\sigma_x/\sigma_z)/\alpha_c \sim 10^{-2}$;
\item the increase of the vertical beam size due to radiation in the
 detector fields must be much smaller than the vertical beam
 size determined by the beam emittance;
\item the non-zero vertical collision angle at the IP due to the
detector field should be corrected;
\item the dilution of the horizontal beam emittance due to the bend 
(needed for obtaining of the required crossing
angle) should be small, and the bending length should be reasonable.
\ei
Below we consider the three last items (the first one is a separate
technical task) and find the maximum value of $\alpha_c$ for
the considered detectors that satisfies all the above criteria.

\section{THE ANGLE BETWEEN TUNNELS, THE BENDING ANGLE}
\begin{figure}[!htb]
\begin{minipage}[b]{0.45\linewidth}
\epsfig{file=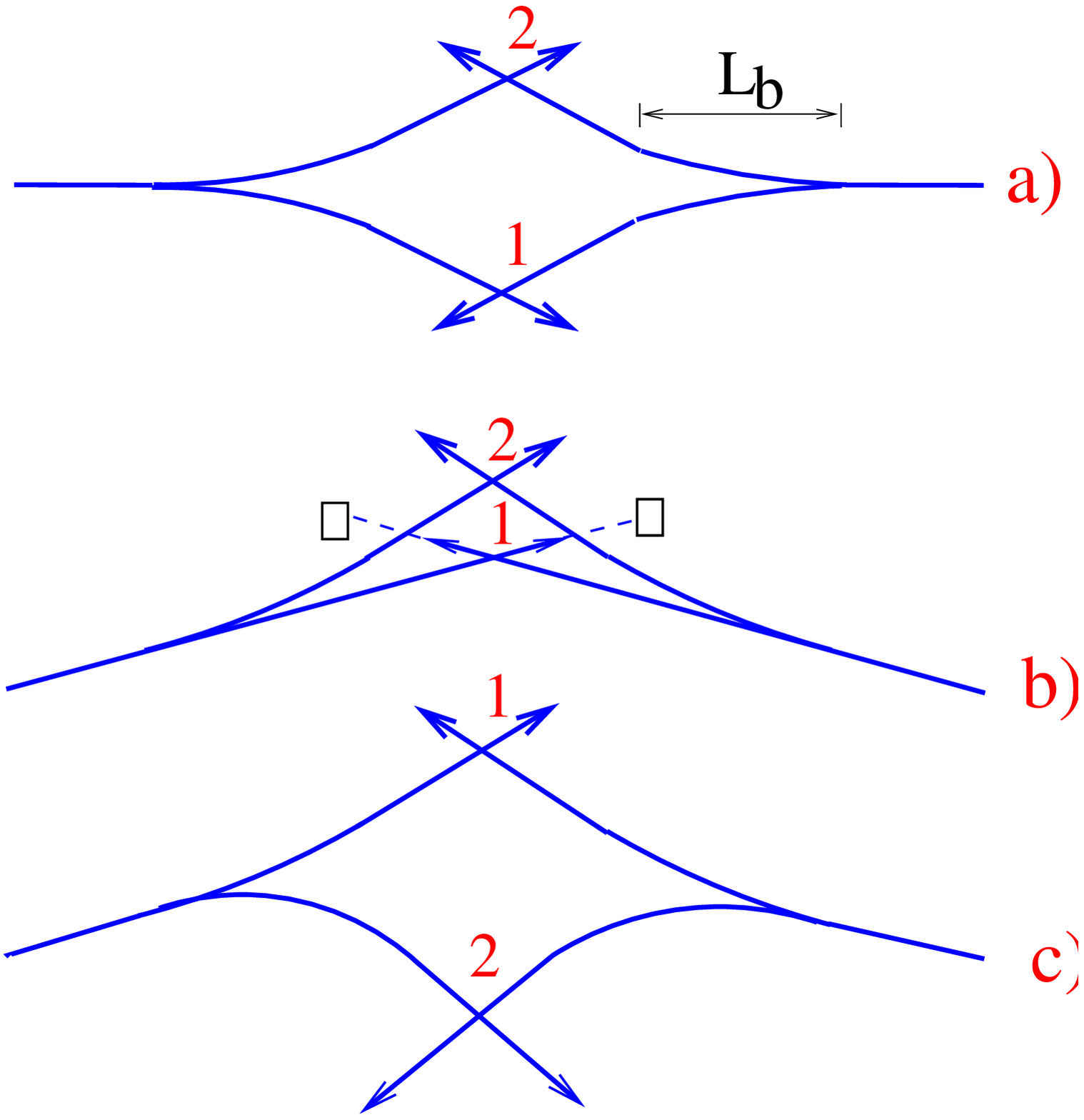,width=8cm,angle=0} 
\end{minipage}
\hspace{-0cm} \begin{minipage}[b]{0.45\linewidth} 
\caption{Possible configurations of tunnels at the ILC. \\ The
crab-crossing angle for \EPEM\ (IP1) is about $\alpha_{c,1}=$ 0--20
mrad, smaller than that for \GG\ (IP2), which is $\alpha_{c,2}\sim$ 25
mrad. \\ [0.3cm] Scheme a: the angle between the tunnels is zero This is
the simplest configuration,  the only problem: for maximum beam
energies the bending length $L_b$ required for a small emittance
dilution may be too long. \\[0.3cm] Scheme b: the angle
between tunnels is non-zero, bending angles are minimum, but there  may be problems with
the space for detectors and for dumping of beams at  IP1. \\[0.3cm]
Scheme c: the angle between tunnels is non-zero, as in the scheme a)
the beams are bent in opposing directions, there is no problem with the
space, but due to different bending angles the maximum energies for IP1
and IP2 are different. This scheme has sense only when tunnels will be
used in future for other multi-TeV linear collider. }
\label{fig2}
\end{minipage}
\end{figure}

Possible configurations of the ILC tunnels with two IPs are shown in
Fig.~\ref{fig2}.
The increase of the normalized horizontal beam emittance due to
synchrotron radiation (SR) after the bending by the angle
$\alpha_b$~\cite{helm,raubenheimer} is
\be \Delta \ENX\ \propto \frac{E^6 \alpha_b^5}{L_b^4}.
\label{emit}
\ee Taking the coefficient from the NLC ZDR~\cite{NLC}, one gets 
\be
\Delta \ENX = 1.8\times 10^{-10}
\left(\frac{2E_0}{\TEV}\right)^6\left(\frac{\mbox{km}}{L_b}\right)^4
\left(\frac{\alpha_b}{10\,\mbox{mrad}}\right)^5 \; \mbox{m}.  \ee 
The bending length corresponding to a 5\% emittance dilution at $\ENX\ =
2\times 10^{-6}$ m and $\alpha_b= 10$ mrad is given in
Table~\ref{tab1} for several beam energies.
\begin{table}[ht]
\caption{The length required for the bending on the angle 10 mrad for
several beam energies} \bc \setlength{\tabcolsep}{3.8mm}
\begin{tabular}{l | c| c| c| c} 
$2E_0$ TeV & 1 & 2 & 3 & 5 \\ \hline $L_b$ km & 0.2 & 0.57 & 1.04 &
2.25 \\
\end{tabular}
\ec 
\label{tab1}
\end{table}

The choice of  scheme depends on the assumed maximum energy of the
 collider in these tunnels. For the ILC with $2E_0=1$ TeV, the required
 bending length is quite small and one can use a zero angle between
 tunnels (scheme a)). For multi-TeV colliders the schemes b) and c) are
 preferable, but in the latter case only one IP can work at the
 highest energy.

\section{MOTION OF PARTICLES IN THE DETECTOR FIELDS} The longitudinal
magnetic field in the detector, $B(z)$, is maximum at the center and
decreases to zero at $z\to \infty$. From conservation of the flux we
can find the radial field (fringe field) at the coordinate $z$ on the beam
trajectory: 
\be B_r= - \frac{\partial B_z}{\partial z} \frac{r}{2} =
-\frac{\partial B_z}{\partial z} \frac{\theta_0 \,z }{2}, 
\label{br}
\ee 
where $\theta_0=\alpha_c/2$. 

The total force acting on the electron in the vertical direction is 
\be 
F_y = e {v \over c} (-B_z\,\theta_0 + B_r)= - e {v \over c} \,\theta_0 \, \left(B_z
+\frac{\partial B_z}{\partial z} \frac{z }{2}\right). 
\label{fy}
\ee 
The vertical coordinate at the position $z$ for a particle traveling
from the infinity is
\be 
y=\int^\infty_z (z^{\prime}-z) y^{\prime\prime}(z^{\prime}) dz^{\prime},
\label{y}
\ee 
where $y^{\prime\prime}\equiv d^2y/dz^2 \approx F_y/P v$.
Substituting (\ref{br},\ref{fy}) to (\ref{y}) and 
integrating by parts, we can calculate the vertical coordinate at $z=0$:
\be y(0)= \frac{e \,\theta_0}{Pc} \left(\int_0^{\infty} B_z \,z \, dz +
\int_0^{\infty} \frac{z^2}{2} dB_z \right) = \frac{e \,\theta_0}{P}
\left(\int_0^{\infty} B_z \,z \, dz + B_z
\,\frac{z^2}{2}\Big|_0^{\infty} - \int_0^{\infty} B_z \,z \, dz
\right) = 0\,.  
\ee 
This interesting result was obtained in ref.\cite{tenenbaum} in a more
complicated way.

The vertical collision angle at the IP is given by 
\be
\theta_y(z=0)=\frac{e\,\theta_0}{Pc}\left(\int^{\infty}_0 B_z dz +
\int^{\infty}_0 {z \over 2} dB_z
\right)=\frac{e\,\theta_0}{Pc}\left(\int^{\infty}_0 B_z dz +B_z{z
\over 2}\Big|_0^{\infty} - \int^{\infty}_0 \frac{B_z}{2} dz
\right)=\frac{e\,\theta_0}{2Pc}\int^{\infty}_0 B_z dz.  
\ee 

We see that the radial fringe field reduces the angle at the IP by a
factor of 2 compared to the action of the longitudinal magnetic
field. This is a new result. 

\begin{figure}[!htb]
\vspace{-0.5cm}
\epsfig{file=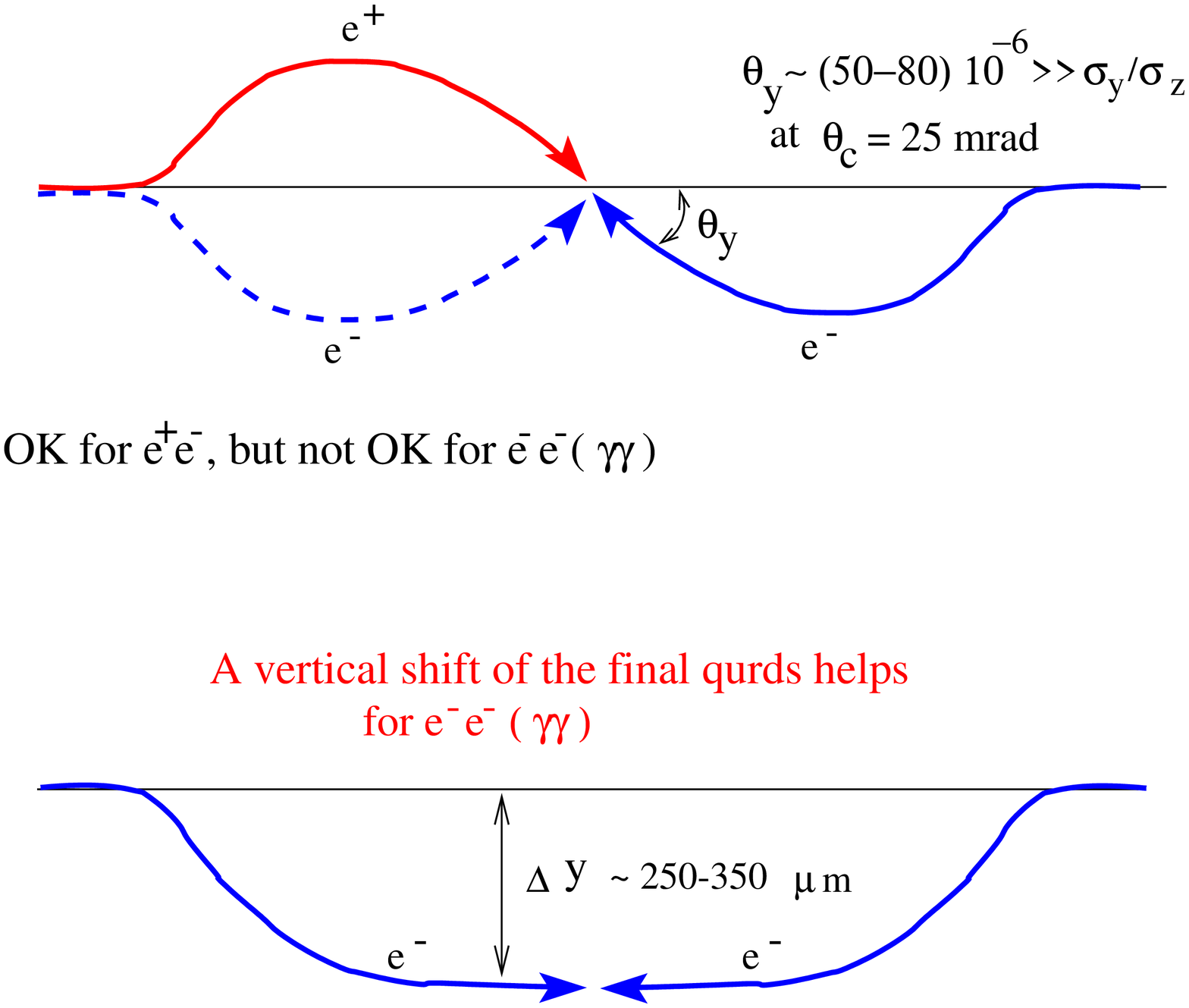,width=10cm,angle=0}
\caption{Trajectories of \EPEM\ and \EMEM\ beams}
\label{fig3}
\end{figure}

It is important to note that in the \EPEM\ case each beam has a
vertical angle at the IP with respect to the horizontal plane but
their angle with respect to one an other is zero: the beams collide
head-on. On the contrary, in the \EMEM\ case  beams collide at the
angle $2\theta_y$, see Fig.~\ref{fig3} (upper).
This situation is acceptable for \EPEM\ collisions. Beams collide
head-on, and the tilt of the collision line is rather small. For
example, the vertical displacement of beamstrahlung photons at the
entrance to the first quadrupole at the distance 4 m from the IP will
only be about 200--300 \MKM. The additional spin-precession angle due
to the tilt of the collision line is of the order of $1^{\circ}$, which is
negligible and can be corrected by the spin rotator at the entrance to
the linac.

  For \EMEM\ collisions, the diagonal angle $\sigma_y/\sigma_z \sim
5\times10^{-7}\,\CM\ /3 \times 10^{-2}\CM\ \sim 1.5\times 10^{-5}$ is
smaller than the relative collision angle $2\theta_y$ by a factor of
ten, which means a considerable loss of  luminosity. The situation
can be corrected by  shift of the first quadrupoles (they will
act as dipole magnets), resulting in the beams  colliding in the horizontal
plane, see Fig.~\ref{fig3} (lower). The vertical displacement of the IP
of 250-350 \MKM\ does not make any problems.

In addition to above method of the  correction there is a suggestion to
use additional coils in the detector (Detector Integrated Dipole), which
allows to correct not only the collision angle but the vertical
position at the IP~\cite{Parker}. I do not think that these
complications are necessary. Moreover, in this case the detector loses
its azimuthal symmetry.

\section{GROWTH OF THE VERTICAL BEAM SIZE DUE TO SYNCHROTRON RADIATION}

The large crab-crossing angle leads to the increase of $\sigma_y$ due to
synchrotron radiation of particles in the detector field. After emission of
the photon(s) the electron comes to the IP with some vertical
deflection.  This effect leads to the decrease of the luminosity. The additional
vertical r.m.s. spread is ~\cite{NLC,TELsnow01}
\be
\sigma^2_{y,SR} = \frac{55 r_e^2}{480\sqrt{3} \alpha}
\left(\frac{e B_s \alpha_c L}{2mc^2}\right)^5. 
\label{sr}
\ee
The number of photons emitted by the electron in the transverse magnetic field
$B$ on the length $L$ is \cite{jackson}
\begin{equation}
N_{\gamma} =\frac{5\alpha e B L}{2 \sqrt{3} m c^2} \sim 0.01 \gamma
\theta_b = 0.005 \frac{eB_s\alpha_c L}{mc^2}\,, \ee where
$\alpha \approx 1/137$ and $\theta_b = eBL/E_0$ is the bending angle. For
example, $B_s=4$ T, $\alpha_c=30$ mrad, $L=4$ m $\Rightarrow$
$N_{\gamma} \sim 1.4$. So, the number of SR photons/electron
$N_{\gamma}\propto L B_s \alpha_c = {\cal{O}}(1)$, which means that the
distribution of $ y$ is non-Gaussian and so one cannot use
Eq.~\ref{sr}. In addition, the fields in detectors are nonuniform. All this
can be accounted for only by simulation.

  The simulation was done using PHOCOL code~\cite{TEL95,TESLATDR},
which simulates beam collisions at linear colliders in all modes.  It
was used for simulation of photon colliders for NLC ZDR, TESLA CDR,
TESLA TDR.  The effect of SR was taken into account in the following way. Using
the map of the magnetic field in the detector (as we saw, $B(z,0,0)$ is
sufficient), the emission of SR photons was simulated and the deviation
of the electron position at the IP from the case when there is no SR
was calculated for each electron.  This deviation was added to the
initial Gaussian position of the electron at $z=0$. Then all particles
were shifted according to their coordinates and angles at the IP to
their starting positions at $z > 5 \sigma_z$ (or CP-IP distance at the
photon collider) and the simulation with account of all collision
effects was started.  In the given simulation (\EPEM\ at $2E_0$=1 TeV)
only attraction between particles was switched on (other processes are
not essential).

 Beam parameters  were taken from U.S. Linear Collider Technology
Option Study: $2E_0=1$ TeV, $N=2\times 10^{10}$, $\sigma_z=0.3$ mm,
\ENX=$9.6\times 10^{-6}$~m, \ENY=$0.04\times 10^{-6}$ m,
$\beta_x=24.4$ mm, $\beta_y=0.4$ mm, $\sigma_x=490$~nm,
$\sigma_y=4$~nm.  

For the \GG\ case, instead of the \GG\ luminosity I simulated the \EMEM\
luminosity (without the $e\to\gamma$ conversion) with
$\sigma_y(\GG)=\sqrt{2}\sigma_y(\EPEM)$ in order to take into account an
effective increase of the vertical beam size due to the ÞÙCompton
scattering. All interactions between particles were switched off. The
position of the first quad (shifted in the \EMEM(\GG) case in order
to have a zero collision angle) was $z=3.5$--$5.7$ m for all detectors.

 The detector fields $B(z,0,0)$ in the LD(TESLA)~\cite{bernard},
SID~\cite{seryi} and GLD~\cite{yamaoka} detectors used for simulation
are presented in Fig.~\ref{fig4}.
\begin{figure}[!htb]
\vspace{-1cm}
\epsfig{file=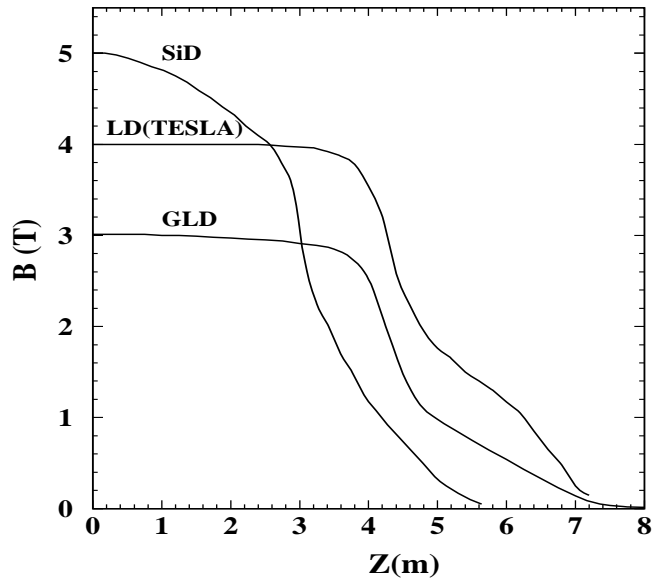,width=10cm,height=10cm,angle=0}
\vspace{-1.5cm}
\caption{T $B(z,0,0)$ in LD, SID and GLD detectors}
\label{fig4}
\end{figure} 

Results of the simulation are presented in Table~\ref{tab2}, the
statistical accuracy is about $\pm$0.5\%. 
\begin{table}[ht]
\bc
\caption{ Results on  $L(\alpha_c)/L(0)$.}  \setlength{\tabcolsep}{3.8mm}
 {  \EPEM\ collisions} \\[4mm]
\begin{tabular}{lllllll}
$\alpha_c$(mrad) & 0 & 20 & 25 & 30 & 35 & 40 \\ \hline
LD(TESLA) & 1. &0.98 &0.95 &0.88 &0.83 & 0.76 \\
SID & 1. &0.995 &0.985 &0.98 &0.95 & 0.91 \\
GLD & 1. &0.995 &0.98 &0.97 &0.94  & 0.925 \\
\end{tabular} \\[0.5cm]
{ \GG\ collisions} \\[4mm]
\begin{tabular}{lllllll}
$\alpha_c$(mrad) & 0 & 20 & 25 & 30 & 35 & 40 \\ \hline
LD(TESLA) & 1 &0.99 &0.96 &0.925 &0.86 & 0.79 \\
SID & 1 &0.99 &0.975 &0.955 &0.91 & 0.86 \\
GLD & 1 &0.995 &0.985 &0.98 &0.97 & 0.93
\end{tabular}\\[2mm]
\ec
\label{tab2}
\end{table} 
We see that the crab-crossing angle $\alpha_c=25$ mrad is acceptable
for all detectors in the \EPEM\ and \GG\ modes of operation.  For
$\alpha_c=30$ mrad, the luminosity loss with the LD(TESLA) detector is already
about 12\%, but possibly it can be decreased by proper shaping of the
magnetic field.

\section{CONCLUSION}

 One of the interaction regions at the linear collider intended for
\EPEM\ and then for \GG\ collisions should have a crab-crossing
angle about 25 mrad.  In this paper I considered problems caused by
such a large crossing angle. The vertical collision angle between
electron beams due to the detector magnetic field can be corrected by
shifting the first quadruples. It is shown that the decrease of the \EPEM\
and \GG\ luminosities due to synchrotron radiation is small up to the
crab-crossing angle $\alpha_c =25$ mrad.  So, one can have the
interaction region at the ILC compatible with both \EPEM\ and \GG\
modes of operation.

\begin{acknowledgments}
I would like to thank A.~Seryi for useful discussions. 

\end{acknowledgments}

\end{document}